\documentclass[nonacm, manuscript]{acmart}

\usepackage{subcaption}
\usepackage{graphicx}
\usepackage{array}
\usepackage{makecell}
\usepackage{xcolor} 
\usepackage{pdfpages}
\usepackage{etoolbox}
\usepackage{booktabs}
\usepackage{multicol}
\usepackage{natbib} 
\usepackage{multirow}
\usepackage{siunitx}
\usepackage{titlesec}
\usepackage{float}
\usepackage{sidecap}
\usepackage{booktabs}
\usepackage{ccicons}          
\usepackage{csquotes}
\usepackage{caption}
\usepackage{fullpage}
\usepackage{enumitem}
\usepackage{makecell}
\usepackage{xpatch}

\makeatletter
\let\@authorsaddresses\@empty
\xpatchcmd{\ps@firstpagestyle}{Manuscript submitted to ACM}{}{\typeout{First patch succeeded}}{\typeout{first patch failed}}
\xpatchcmd{\ps@standardpagestyle}{Manuscript submitted to ACM}{}{\typeout{Second patch succeeded}}{\typeout{Second patch failed}}    \@ACM@manuscriptfalse
\makeatother

\renewcommand\footnotetextcopyrightpermission[1]{} 
\setcopyright{none}
\title{Anatomy of a Rumour: Social media and the suicide of Sushant Singh Rajput}

\author{Syeda Zainab Akbar}
\affiliation{
  \institution{Microsoft Research$^{*}$}
   \country{India}
   \thanks{$^{*}$The views contained in this research paper are those of the authors’ and do not reflect the views and opinions of Microsoft.}}
 
 \author{Ankur Sharma}
\affiliation{
   \institution{Microsoft Research}
  \country{India}}
 
 \author{Himani Negi}
 \affiliation{
   \institution{Microsoft Research}
   \country{India}}
 
 \author{Anmol Panda$^{+}$}
 \affiliation{
 \institution{Microsoft Research}
   \country{India}
   \thanks{$^{+}$At the time of publication, Anmol Panda was associated with the University of Michigan. Some of the work intersects with his time as a student there.}}
   
\author{Joyojeet Pal}
 \affiliation{
 \institution{Microsoft Research}
   \country{India}}
 
\begin{abstract}
 The suicide of Indian actor Sushant Singh Rajput in the midst of the COVID-19 lockdown triggered a media frenzy of prime time coverage that lasted several months and became a political hot button issue. Using data from Twitter, YouTube, and an archive of debunked misinformation stories, we found two important patterns. First, that retweet rates on Twitter clearly suggest that commentators benefited from talking about the case, which got higher  engagement than other topics. Second, that politicians, in particular, were instrumental in changing the course of the discourse by referring to the case as `murder', rather than `suicide'. In conclusion, we consider the effects of Rajput's outsider status as a small-town implant in the film industry within the broader narrative of systemic injustice, as well as the gendered aspects of mob justice that have taken aim at his former partner in the months since.

\end{abstract}

\keywords{Sushant Singh Rajput, SSR, Rhea Chakraborty, CBI, Suicide, Depression, Social Media, Misinformation, Bollywood, Twitter, YouTube, India}

\renewcommand{\shortauthors}{}
\renewcommand{\shorttitle}{}
\begin{document}

\maketitle

\section{Introduction \& Background}
On June 14, 2020, Hindi cinema and television actor Sushant Singh Rajput took his own life in his apartment in Mumbai. The 34-year old actor had featured in several  successful and critically acclaimed films and television sitcoms in the past decade, and was among the top-billed stars in the industry at the time of his death. 

His death happened during the COVID-19 lockdown in India, and triggered active discussion on social media -- grief, mental and emotional health during the lockdown, and the film industry in general. In the weeks to follow, the story and several threads emanating from it continued to grab a significant mindshare, on both social media, and mainstream news.

In this study, we map the timeline of the coverage related to the case, both through mainstream channels and on social media feeds of media houses, journalists, and politicians in India. The story and its evolution online is instructive in understanding how the affective contours of a tragedy can be turned into a media event, and in turn, how that can develop a life and trajectory of its own. The case offers an insight into the ability of media influencers to steer the trajectory of a story, but also a warning about the after-lives of pseudo events.
\section{Data and Methodology}
We analyzed the data from three sources - the YouTube pages of mainstream television news channels, Twitter trending hashtags, and Tweets from politicians, influencers, journalists, and media houses in India, and an archive of debunked misinformation stories compiled from factcheckers operational in India. 

We analyzed these, in addition to which we created a visualization of all the key events and dates, the news coverage on those dates, and the misinformation about the case or its key related persons that was debunked on those dates by factcheckers.\footnote[1]{http://joyojeet.people.si.umich.edu/ssr/}. Due to the size of the visualization, it is not included in this document.

\paragraph{YouTube pages of Television News:}
We studied the YouTube pages of the 5 most viewed English (India Today, CNN-News18, Republic World, Times Now, DD News) and Hindi (Aaj Tak, ABP News, Republic Bharat, News 18 India, Zee news) television news channels in India. 

We used the YouTube data API v3 to extract 7171 videos from the news channels’ YouTube handle. Each of these channels had dedicated playlists for Sushant Singh Rajput’s suicide except for Zee News and DD news for which we viewed all publicly uploaded videos between 14th June and 12th September and manually annotated them for those relevant to the case based on their content. 

Since these videos are hosted online, their comparative appeal may not be representative of these media houses' viewership. Nonetheless, they serve as a proxy, especially to examine the engagement these media channels get, as against other programming content.

We used the tags used by the news channels on their videos to do a temporal topical analysis. Since the purpose of these tags is to improve search engine optimization, these terms may be skewed towards getting more hits than to reflect the full extent of the linked story. We removed all the redundant words (channel names, times, etc.) to create word clouds of topics.

\paragraph{Twitter Trending Topics:}
We maintained a repository of hourly trending hashtags on Twitter in India since June 14, 2020, which can be obtained using the Twitter Public API. These topics were manually annotated by two researchers for a simple binary of whether a topic is related to Sushant Singh Rajput for the period of June 14th and September 12th, 2020. The inter-coder reliability score using Cohen’s Kappa gave a score of 0.876. The rest of the hashtags were arbitrated and re-annotated for disagreements.

\paragraph{For Twitter Media Influencer Data:}
On Twitter, we extracted tweets from a list of over 2000 journalists and media houses and 1200 politicians in two states where the issue has featured significantly on the news - Maharashtra, where Rajput lived and worked, and Bihar, his place of origin. We used the dataset from Nivaduck \cite{nivaduck}, an open source state-party annotated dataset of more than 36000 politicians in India at various levels of different regional and national parties. We have made the lists of accounts available.\footnote[1]{http://joyojeet.people.si.umich.edu/ssr/}.

For media houses and journalists, we used the annotated list of media houses from our dataset of influencers in India. Using these datasets, we pulled 103125 tweets of 7818 politicians- specifically, 4172 tweets of 274 Bihar politicians and 17328 tweets from 980 Maharashtra politicians, 45969 tweets of 239 media houses and 41056 tweets from 1930 journalists who actively tweeted between June 14th and September 12th on Sushant Singh Rajput related topics.

The determination of whether a topic is related to Sushant Singh Rajput was done by human annotators familiar with the news cycle. For analyzing the tweet text, we used standard natural language processing data preparation techniques including lower casing, removing stop words, tokenizing, etc. We also removed terms such as `breaking', `latest', `primetime' etc. which do not have contextual value.

We maintained a sliding dictionary of hashtags, with a window size of 3 days, to create temporal graphs, and only considered the new hashtags that were not present in the dictionary then. We did this to weigh in favour of new terms or hashtags that trended, reducing the impact of repetitive threads that had been around for a while. The window size was chosen after observing multiple rounds of many event stories around Bihar, Maharashtra, Media Houses, etc and had been kept same for all.

This paper is not hypothesis driven. Our goal is to offer a preliminary survey of the lay of the land for more empirically-driven results in the future.
\section{Findings}
We found that the various actors - politicians, media houses, and journalists framed their own narratives in relation to the incident. These topical preferences reflect individual biases, perceptions, motivations and gains.

\subsection{Descriptive term and handle-frequency analysis of various actors in social media}

In this section, we examine the content of key actors' tweets, the temporal patterns around when and how often they discuss specific subjects, and how their audiences react to these.

\subsubsection{Twitter Actors: Politicians, Media Houses and Journalists:}

To set the tone of what comprises and what differentiates the key actors on the Sushant Singh Rajput's discussion, we take a look at basic word clouds of what their tweets talk about. 

\begin{figure}[ht]
    \centering
    \includegraphics[width=1.0\linewidth]{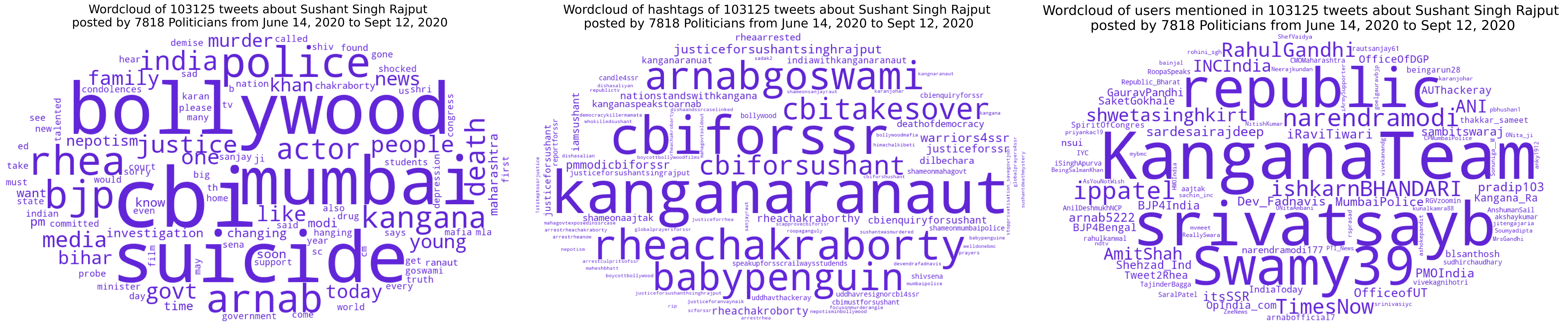}
    \caption{Wordclouds of all Tweets, Hashtags and Mentions from all the politicians in India}
    \label{fig:overall-stuff}
\end{figure}

\noindent \textbf{All politicians}

\begin{figure}[ht]
    \centering
    \includegraphics[width=0.5\linewidth]{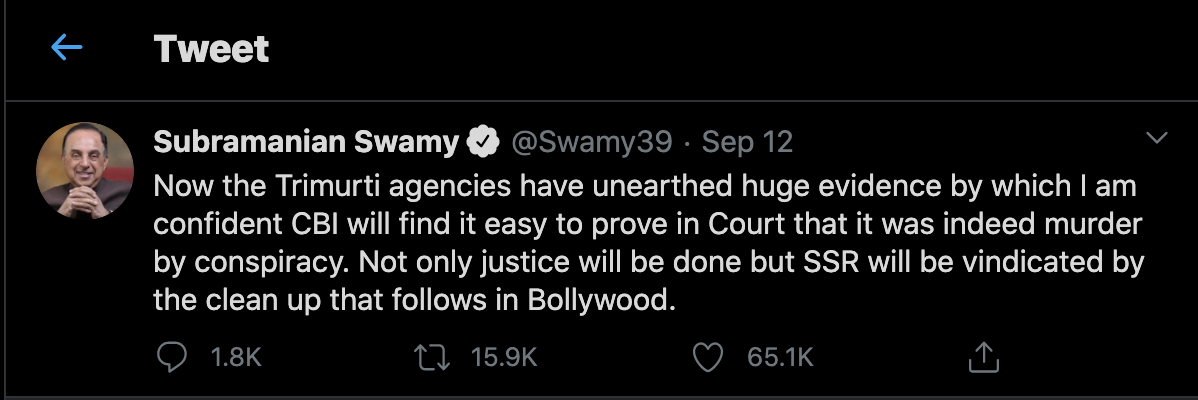}
    \caption{An example of a tweet with unverified innuendo from Subramanian Swamy on Sept 12th, 2020}
    \label{fig:swamy}
\end{figure}

In the figure \ref{fig:overall-stuff}, we see what politicians have been talking about in wordclouds during the study period. 7818 politicians from across various parties, mentioned text or used hashtags in a total of 103125 tweets during the study period. The left, and center wordclouds show the spread of words in tweets and hashtags respectively, while the wordcloud on the right is a list of the most used handles, either through retweets or direct mentions. Here are a few key patterns:

\begin{itemize}
\item \textbf{Words in Tweet text:} The extensive use of `Rhea Chakraborty', the former partner of the deceased, the references to the `Central Bureau of Investigation (CBI)' and two commentators, `Arnab Goswami' and `Kangana Ranaut', who have made incendiary remarks on the case show politicians' engagement in speculation. 

    \item \textbf{Top hashtags:} The use of hashtags by politicians shows a systematic targeting of the Shiv Sena (\#UddhavResignOrCBI4SSR, \#ShameOnMahaGovt, \#BabyPenguin), the last being a disparaging reference aimed at Aditya Thackeray. This shows that the hashtag usage by politicians is dominated by those inimical to the present Maharashtra government, largely the BJP. While key people involved in the discourse such as Goswami and Ranaut feature here as well, we also see a number of self-styled justice hashtags including \#Warriors4SSR and \#IAmSushant. The hashtags are also used to highlight a lack of faith in local law enforcement's ability to manage the case.
    
    \item \textbf{Top mentions:} The most featured individuals include politicians across parties. While Narendra Modi (@narendramodi) and Amit Shah's handles (@AmitShah) are mentioned, they are largely callouts by other politicians with requests to intervene. Likewise, Subramanian Swamy's account (@Swamy39) is called out, since he has wide reach, and given that he has engaged with incendiary misinformation in the past (figure \ref{fig:swamy}). The accounts engaging with the subject are from across parties - for instance YB Srivatsa (@srivatsayb) is from the INC, while Ravi Tiwari (@iRaviTiwari) is with the BJP. The key engagement with the subjects is through callouts to or tweets from handles like @KanganaTeam, @ishkarnBhandari, and @IPPatel, all of which have been known to put forth provocative, unverified information. 

\end{itemize}

\noindent \textbf{Journalists and Media Houses}

\begin{figure}[ht]
    \centering
    \includegraphics[width=1.0\linewidth]{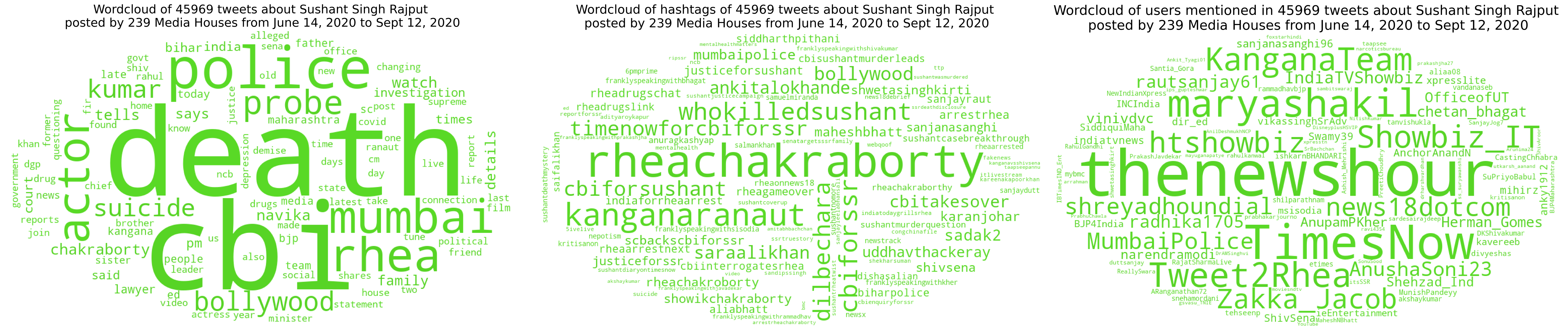}
    \setlength{\abovecaptionskip}{-0.1cm}
    \caption{Tweets, Hashtags and Mentions from Media Houses in India}
    \label{fig:media_houses_overall}
\end{figure}

In figure \ref{fig:journos_overall} below, we see what journalists have been talking about in wordclouds during the study period. 1930 journalists were sampled \footnote[2]{http://joyojeet.people.si.umich.edu/ssr/}. To the left, we see the hashtags, the center figure shows the mentions of individual accounts, and the figure to the right is a wordcloud of the most used terms in the Tweets. In figure \ref{fig:media_houses_overall}, we see the wordclouds for media houses.

\begin{figure}[ht]
    \centering
    \includegraphics[width=1.0\linewidth]{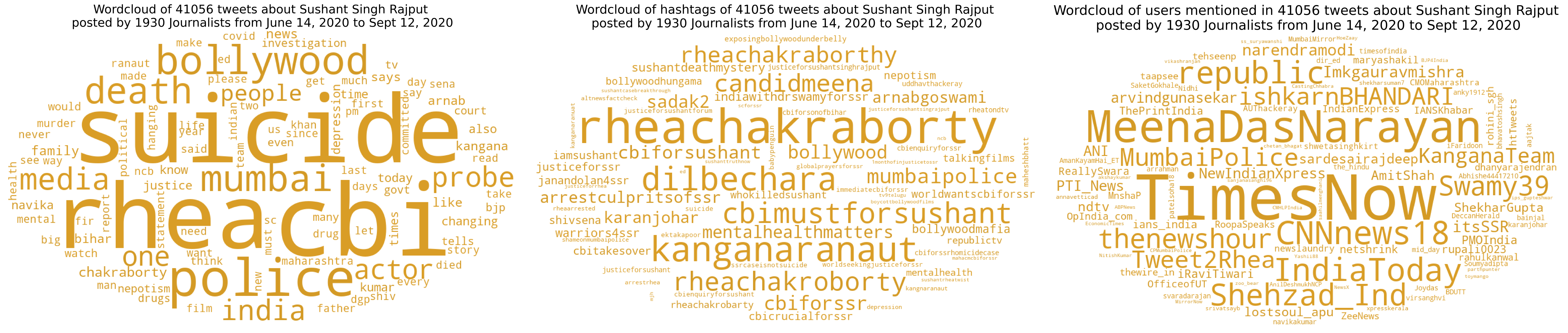}
    \setlength{\abovecaptionskip}{-0.1cm}
    \caption{Tweets, Hashtags and Mentions from Journalists in India}
    \label{fig:journos_overall}
\end{figure}

\begin{itemize}

    \item \textbf{Words in Tweet text:} The emphasis of journalists on the `Mumbai police' and `Bollywood' digresses from the individual narratives that focus on Sushant Singh Rajput's reasons for committing suicide.

    \item \textbf{Hashtags:} 
   Journalists and media houses seem to be equally complicit in pushing an agenda against Rhea Chakraborty (\#RheaChakraborty), whose name was the most used hashtag by both individual journalists and the mainstream media. Both also pushed for the CBI inquiry (\#CBIMustForSushant and \#CBIForSSR). We also see that media craft themselves as vigilantes (\#TimesNowForCBIForSSR and \#Warrior4SSR) and present the argument that there is a cause for doubt in the minds of audiences (\#ArrestCulpritsofSSR and \#SushantDeathMystery). We also see hashtags that seek to target key figures in the Hindi film industry including Karan Johar, and the film, Sadak 2.

    \item \textbf{Top mentions:} 
    The most mentioned individual Twitter handles show an intersection with the more retweeted figures by politicians - including @KanganaTeam @swamy39 who have made public statements or presented innuendo on the case, as well as lesser known individuals including @ishkarnBhandari and @MeenaDasNarayan who came to public prominence as a result of their claims on the case.  News channels such as @TimesNow @CNNnews18 and @republic channel prominently featured in the messaging related to the story, while few specific shows including @thenewshour trended due to their significant coverage.

\end{itemize}

\subsubsection{YouTube: Television Channels}

We examined the content and meta data of the content from 10 television news channels posted on YouTube, and found that a number of tactics are employed to keep the story newsworthy. Some of the trends we see are as follows:

\begin{itemize}
\item The term suicide was used the highest in the first week after his death and was also strongly visible in the first month, but the term started to get less common in the second month and continued to decrease as a dominant subject.

\item The discussion on Bollywood nepotism began early on, positioning Rajput as an outsider. By July, conversations on specific persons and professional rivalries started to become more common - director Sanjay Leela Bhansali and Reshma Shetty made the news. 

\item The filing of a First Information Report (FIR) in Bihar drew news coverage since a police team from Bihar arrived in Mumbai and had a run in with the Mumbai police. The angles vilifying the Mumbai police, specifically presenting Rajput as an outsider to the industry, and to a hostile city with a cabalish industry coalesced. 

\item As the Bihar police investigation dead-ended, calls for a CBI investigation intensified across channels. By mid-July, Rhea Chakraborty became the main topic of discussion on television channels' YouTube content.

\item Periodically, new characters were introduced, including close friends Ankita Lokhande and Siddharth Pitani. His former cook and longtime friend Sandip Singh emerged in the second month. New characters brought their own threads to conversations, consequently generating new interest.

\item The third month of television coverage focused on the angle of drugs, both following Rhea Chakraborty's arrest and a series of new stories which again have developed their own trajectories independent of the inital story around Rajput. 

\end{itemize}

\subsection{Temporal analysis of political and media actors on social media}

\subsubsection{Discourse drivers on Twitter}

\begin{figure}[ht]
    \centering
    \includegraphics[width=1.0\linewidth]{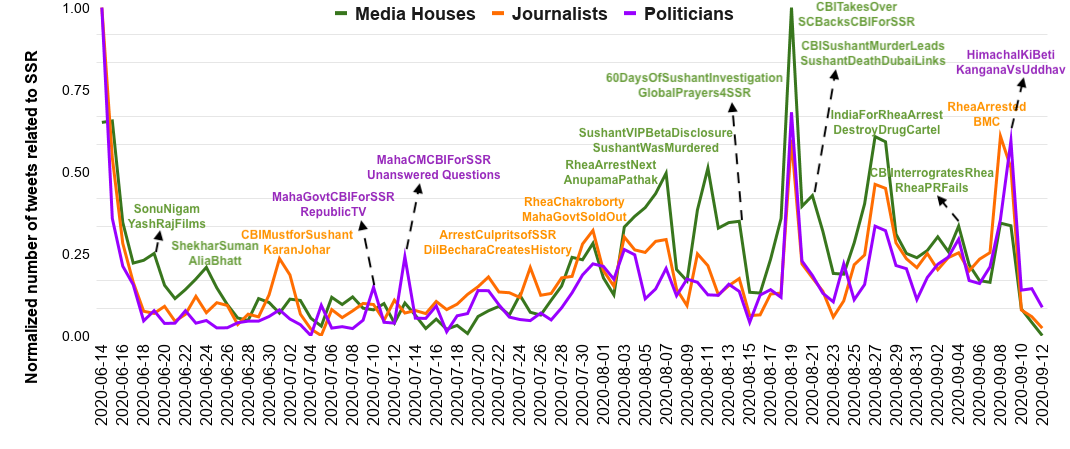}
    \setlength{\abovecaptionskip}{-0.1cm}
    \caption{Normalised tweet counts for all politicians, media houses and journalists}
    \label{fig:media_journo_normalised}
\end{figure}

Figure \ref{fig:media_journo_normalised} has a timeline of the volume and peaks of major events on Twitter. While we don't have a sense of what is happening on largely closed or encrypted platforms like Facebook or WhatsApp, the timeline of peaks offers an indicator of when the story became newsworthy and how its purchase grew. Some key patterns seen are as follows: 

\begin{itemize}

        \item At different points, politicians or media houses and journalists have caused spikes in the amount of Twitter traffic on the Sushant Singh Rajput case. As the figure shows, there is an overlap in the peaks between when politicians talked about the subject, and when media persons did.
        
        \item There was one point when politicians were more active in pushing the Twitter traffic. This happened during mid-July, when politicians started a coordinated effort to ask for a CBI enquiry. The majority of accounts active there belonged to BJP politicians. Interestingly, journalists also pushed a strong anti-Maharashtra government narrative in early August 2020.
        
        \item The first point when journalists drove the Twitter discourse was a large volume of tweets targeting Karan Johar, which circulated two weeks after the suicide.
        
        \item New dramatis personae have trended roughly every two weeks, starting with Sonu Nigam and Yash Raj Films, followed by Shekhar Suman, Alia Bhatt, Karan Johar, Rhea Chakraborty, Anupama Pathak, Mahesh Bhatt, Ankita Lokhande, Kangana Ranaut, Disha Salian, Aditya Pancholi etc. Usually the pattern is for one name to emerge and then disappear over time. 
        
        \item The media houses replaced individual journalists as the key drivers of the discourse by late July. A significant point when individual media houses, specifically prime time news, managed to corner the narrative was around demands for Rhea Chakraborty's arrest in early August. 
        
        \item News channels also drove the idea of murder around early August, following which the Twitter traffic increased dramatically, peaking at the handover of the case to the CBI in mid-August.

\end{itemize}

\subsubsection{Content posted on TV channels with view counts}

\begin{figure}[ht]
    \centering
    \includegraphics[width=1.0\linewidth]{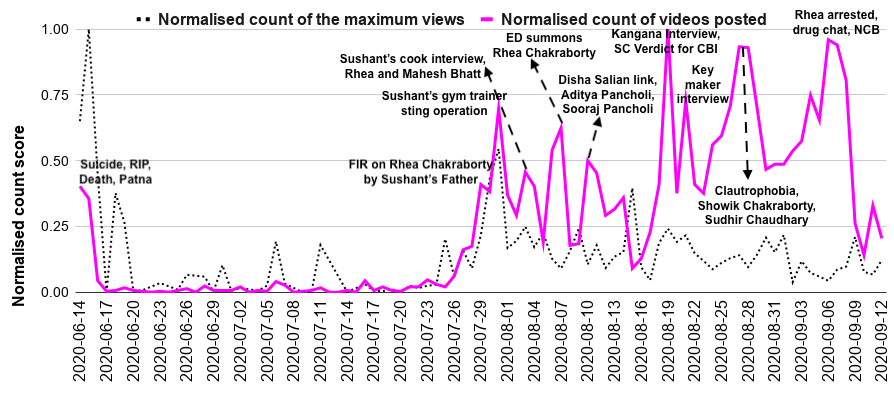}
    \setlength{\abovecaptionskip}{-0.1cm}
    \caption{Normalised view count and video count posted on television}
    \label{fig:tv_topical}
\end{figure}

Figure \ref{fig:tv_topical} has a timeline of the volume and peaks of major events on YouTube. We see an important difference between what happened on Twitter and what happened on YouTube. While on Twitter, there was a fairly consistent amount of activity in late June and early July, the YouTube activity took off much more aggressively in the last week of July following the anti-Maharashtra government activity on Twitter and the filing of an FIR against Rhea Chakraborty.

\begin{itemize}
       \item{TV channels took a much more aggressive stance towards finding characters, possibly because the medium relies on visual contact. Unlike on Twitter, where the focus was on highly speculative conversations about celebrities, TV channels focused on those they could find and interview.}
        \item{Viewers' count rose sharply with key interviews with various people who had been in contact with Rajput, including his trainer, cook, key maker etc.}
        \item{Speculative material, including the linking of the case with Disha Salian's death found high viewership online.}
        \item{There was a peak in viewership around the interview of Kangana Ranaut.}
        \item{There was a second peak during the arrest of Rhea Chakraborty, but a fall immediately after. The traffic then segued into stories about drug use among celebrities.}
    \end{itemize}

\subsubsection{Frequency of tweeting/posting by various actors}

In Table 1, we see the overall traffic and the source of tweets and videos from various actors over time.  Overall, we see that the amount of online traffic has increased significantly across Twitter and YouTube after the first month since the actor's death.

\begin{table}[ht]
\centering
\caption{Frequency table depicting monthly number of tweets/videos for various actors from 14th June 2020 to 13th September 2020}
\label{tab:freq_table}
\begin{tabular}{c|c|c|c|c|c}
\toprule
\textbf{Source} & \multicolumn{2}{c|}{\textbf{Actors}} & \textbf{14th June - 13th July} & \textbf{14th July - 13th Aug} & \textbf{14th Aug - 13th Sept} \\ \midrule
\multirow{5}{*}{\textbf{Twitter}} & \multirow{3}{*}{\textbf{Politicians}} & \textbf{Bihar} &  809 & 1836  & 1527 \\ \cline{3-6} 
 &  & \textbf{Maharashtra} &  3564 & 5508 & 8256 \\ \cline{3-6} 
 &  & \textbf{All} &  26183 &  30460 & 46482 \\ \cline{2-6} 
 & \multicolumn{2}{c|}{\textbf{Journalists}} &  11217 & 13637 & 16202 \\ \cline{2-6} 
 & \multicolumn{2}{c|}{\textbf{Media Houses}} &  12118 & 14374 & 19477 \\ \hline
\textbf{YouTube} & \multicolumn{2}{c|}{\textbf{TV Channels}} &  307 &  1965 & 4899 \\ \bottomrule
\end{tabular}
\end{table}

\begin{itemize}
    \item In terms of the overall volume of social media traffic, we see that there was an initial peak of interest in Bihar in July, first around the wrangling of the case between Maharashtra and Bihar police, and then on the overall question of how people from Bihar are treated in Mumbai.
    \item Initially, it seemed as though Sushant Singh Rajput would become a major case for the Bihar elections, however, after peaking in July, the number of politicians in Bihar talking about the case fell.
    \item On the other hand, the case has continued to be very significant among Maharashtra politicians, and has grown in importance among politicians around the country.
    \item The media houses and journalists have seen a consistent increase in the amount of social media discussion on the subject.
    \item The most significant media output growth has been on YouTube, where television channels have increased their uploads tenfold from the time of the suicide. This suggests a corresponding increase in the time devoted to the case on live television news programming as well.  
    
\end{itemize}

\subsubsection{Party-wise Comparison: BJP vs INC}
We see a few differences in the ways that the national ruling and opposition parties respectively discuss and frame the issue. In Maharashtra, however, the BJP is in the opposition, and the politicians can take a more aggressive anti-government, anti-police stance.

\begin{figure}[ht]
    \centering
    \includegraphics[width=1.0\linewidth]{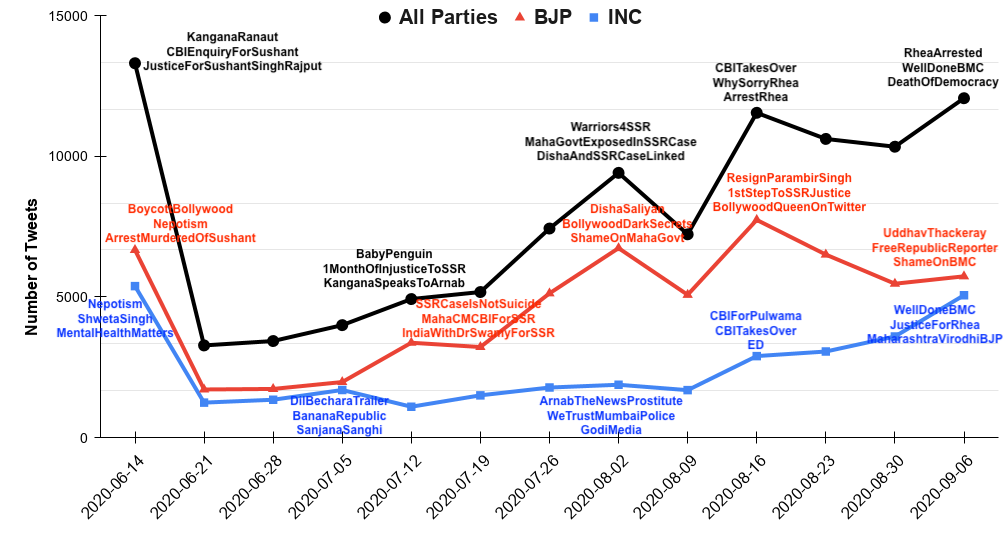}
    \setlength{\abovecaptionskip}{-0.1cm}
    \caption{Weekly comparison of number of tweets related to SSR from different parties.}
    \label{fig:bjp_inc_trend}
\end{figure}

\begin{itemize}
    \item First, we find that although both parties have roughly the same number of known politicians talking about the subject (3037 INC politicians versus 3478 BJP politicians), we find that the BJP is almost twice as voluminous in its tweeting (61196 tweets compared to INC's 32406) 
    \item INC uses the term `suicide' much more than the BJP. BJP uses `Bollywood' a lot more than the INC. BJP's general approach is to use the case to discuss Bollywood in negative light. 
    \item Television anchor `Arnab Goswami' is the second most used term besides `Sushant Singh' and `suicide', as well as the most used hashtag by INC. In contrast, BJP very sparsely discusses Arnab Goswami. Mostly, INC leaders refer to Arnab Goswami in a negative context using hashtags like \#ArnabTheNewsProstitute.
    \item `Kangana Ranaut' is a frequently used hashtag by both parties, but the messaging is generally pro-Kangana on the BJP side with hashtags like \#NationStandsWithKangana, but negative on the INC side, which uses hashtags like \#CastesistKangana to refer to her.
    \item The top hashtags of BJP are around invoking the CBI such as \#CBIForSSR, but also a significant number are direct attacks at the Maharashtra government such as \#BabyPenguin and \#MahaGovtExposed
    \item The top mentions on the BJP side are for Twitter accounts with major online social media following including Subramanian Swamy, Kangana Ranaut and Republic TV. The Congress on the other hand mainly refers to individuals from its own outreach team including YB Srivatsa. The Congress' approach gets less attention online.
    
\end{itemize}

\noindent \textbf{Frequency comparison between BJP and INC.} (Refer to figure \ref{fig:bjp_inc_trend})

\begin{itemize}
    \item BJP dominated the overall trends on Twitter in terms of the total number of tweets for most of the period up until the end of August.
    \item The CBI case declaration has been a major push for the BJP's Twitter footprint.
    \item Towards the end of the study period, around early September 2020, the INC started to approach the same volume of social media engagement as the BJP, partly driven by the \#JusticeForRhea campaign, because by this point Sushant Singh Rajput's partner Rhea Chakraborty had been arrested and lodged in jail without bail.    
\end{itemize}

\subsection{Comparison of topical choices}

\subsubsection{Suicide, Murder, or Mystery?}
We mapped the usage of the terms `suicide', `murder', and `mystery' to examine the ways in which the discourse has evolved over time. We also find some descriptive patterns in how some actors prefer one term to another.

\begin{figure}[ht]
  \begin{subfigure}[b]{0.5\linewidth}
    \centering
    \includegraphics[width=1.0\linewidth]{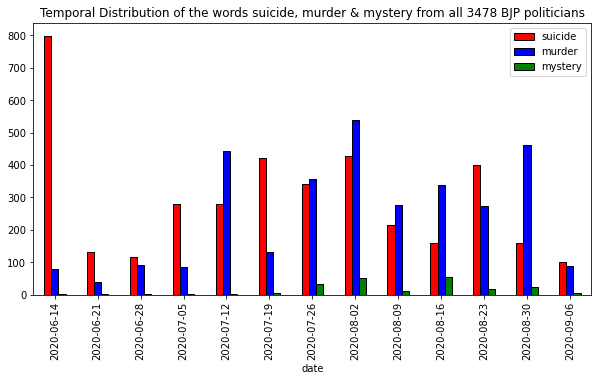}
    \setlength{\abovecaptionskip}{-0.1cm}
    \caption{BJP}
    \label{fig:murder_mystery_bjp}
  \end{subfigure}
  \begin{subfigure}[b]{0.5\linewidth}
    \centering
    \includegraphics[width=1.0\linewidth]{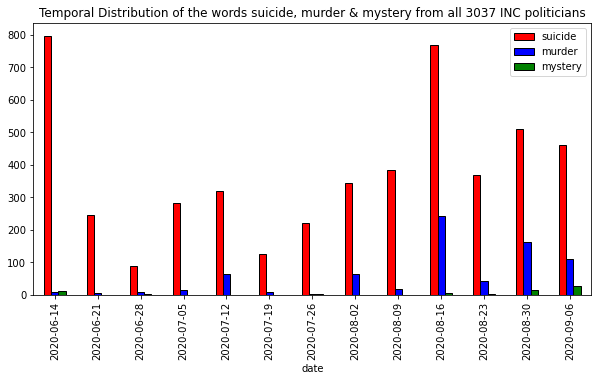}
    \setlength{\abovecaptionskip}{-0.1cm}
    \caption{INC}
    \label{fig:murder_mystery_inc}
  \end{subfigure}
  \caption{Comparison of the tweets containing `Suicide'/`Murder'/`Mystery' amongst politicians }
  \label{fig:murder_mystery_bjp_inc}
\end{figure}

\noindent \textbf{\\Politicians (figure \ref{fig:murder_mystery_bjp_inc})}
\begin{itemize}
    \item The ruling BJP shows a strong discernible pattern of preferring `murder' over `suicide'. We find that the usage of `suicide' keyword dropped sharply right after one week following the actor's death. 
    \item Over the weeks that followed, there was an increased usage of `murder' keyword repeatedly in tweets by BJP politicians, while the opposition INC consistently used the term `suicide' more than `murder', despite a small rise in the use of `murder' in mid-August. 
    \item Overall, the data strongly suggest that the BJP drove the insinuation of `murder' since it was used more than `suicide' in most weeks since July.
    \item Politicians in Bihar were on average much more likely to refer to the case in terms of `murder' than `suicide', in comparison to politicians in Maharashtra.
    
\end{itemize}


\noindent \textbf{Journalists and Media Houses}
\begin{itemize}
    \item For journalists and media houses, there is a sharp dip in the usage of `suicide' narrative after the first week. In absolute numbers, the usage of the `suicide' keyword dropped from 1400 to 350 for journalists, and from 680 to 280 for media houses.
    \item The frequency of `murder' to refer to the death steadily increased over the months that followed for both media houses and journalists.
\end{itemize}

\begin{figure}[ht]
    \centering
    \includegraphics[width=0.5\linewidth]{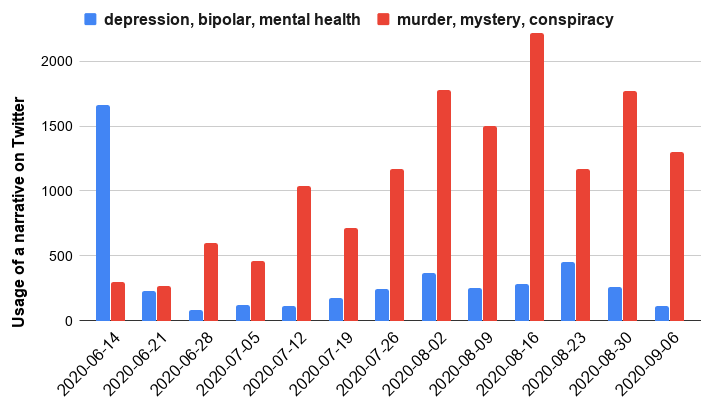}
    \setlength{\abovecaptionskip}{-0.1cm}
    \caption{Comparison of the usage of two different narratives on Twitter by politicians, media houses and journalists}
    \label{fig:depression_murder_narratives}
\end{figure}

We see in the frequency of terms around `depression' and `mental health' as compared to conspiracy related terms above that while `depression' and `mental health' were part of the initial discourse, the conversation has been aggressively captured by discussions on conspiracies and murder (Refer figure \ref{fig:depression_murder_narratives}).

\subsection{Politicians, Police, and Warring States}

The Sushant Singh Rajput case exposed a number of faultlines in the ways Indians from various groups and classes perceive tensions in their relations with other groups. While the story on Rajput's outsider status quickly fed into stories about cabals in the entertainment industry, the angle of him being a migrant to the westernized metropolis of Mumbai from the Gangetic heartland became highlighted. 

The situation escalated when a police complaint was lodged in a station in Bihar, as a result of which an officer needed to be sent to Mumbai to investigate. The head of the police in Bihar went on a widely covered television interview, on which he spoke disparagingly of Rajput's partner. He had more media coverage when the officer sent to Mumbai was unable to do his investigations due to COVID quarantine rule. The officer became a public figure as a result of his appearances and would go on to leave the police to become a politician within a short period of time.

\subsubsection{The Police}

\begin{figure}[ht]
  \begin{subfigure}[b]{0.5\linewidth}
    \centering
    \includegraphics[width=1.0\linewidth]{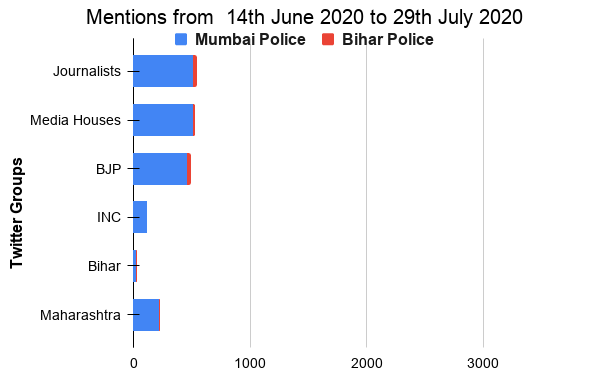}
    \setlength{\abovecaptionskip}{-0.1cm}
    \caption{14th June to 29th July}
    \label{fig:mumbai_bihar_first}
  \end{subfigure}
  \begin{subfigure}[b]{0.5\linewidth}
    \centering
    \includegraphics[width=1.0\linewidth]{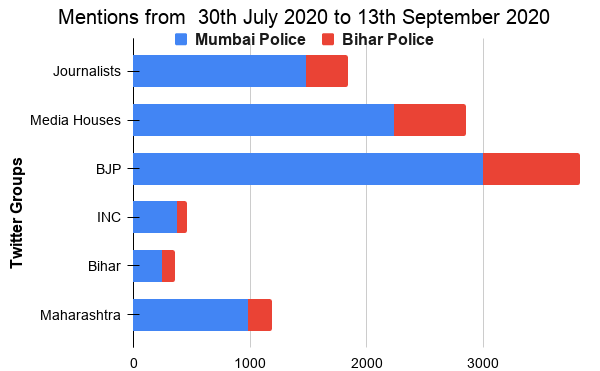}
    \setlength{\abovecaptionskip}{-0.1cm}
    \caption{30th July to 13th September}
    \label{fig:mumbai_bihar_second}
  \end{subfigure}
  \caption{Comparison of the number of tweets containing Mumbai/Bihar police in the two halves}
  \label{fig:mumbai_bihar_total}
\end{figure}


\begin{itemize}
    \item As we see in figures \ref{fig:mumbai_bihar_first} and \ref{fig:mumbai_bihar_second}, Mumbai police has been at the receiving end of several thousand tweets by various stakeholders throughout the entire period, though particularly after late July 2020.
    \item The BJP has been the most active in attacking the city police of Mumbai.
    \item Pro-Maharashtra politicians came out with tweets including \#WeTrustMumbaiPolice, \#WeStandWithMumbaiPolice, while those opposed to the government (including Bihar politicians) used tags including \#ResignParambirSingh (referring to the police commissioner) and,   \#ShameOnMumbaiPolice. The attack on the police was used as a bridge into attacking the state cabinet, beginning with  \#AnilDeshmukhSavingSSRKillers (the home minister), \#MahaGovtExposed, and eventually the Aditya Thackeray, who is already one of the most trolled politicians in India.
    \item The Bihar police did not receive much antagonistic attention from Maharashtra, despite their abortive attempt to play a role in the case.
\end{itemize}

\subsubsection{Bihar and Maharashtra political Twitter timeline}

\begin{figure}[ht]
    \centering
    \includegraphics[width=1.0\linewidth]{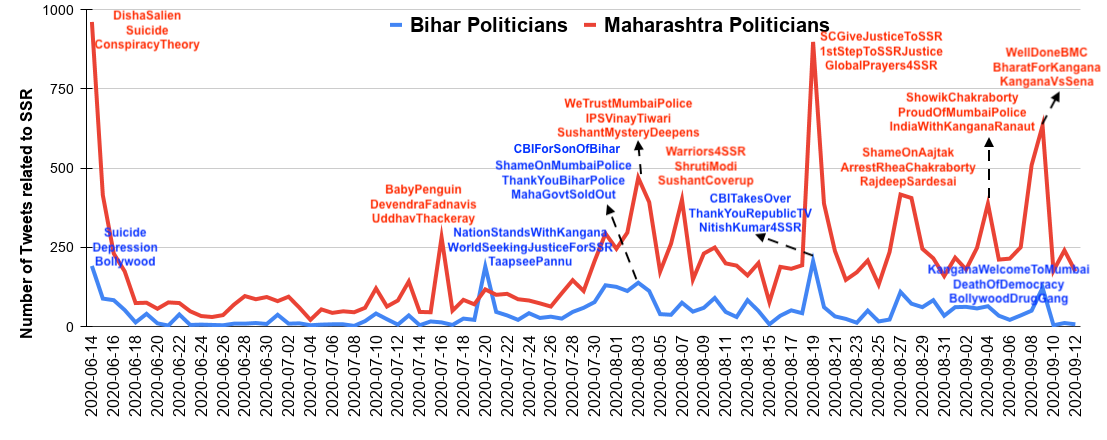}
    \setlength{\abovecaptionskip}{-0.1cm}
    \caption{Tweet counts with labelled peaks for all politicians in Bihar and Maharashtra}
    \label{fig:cout_bihar_maha}
\end{figure}

\begin{itemize}
    \item Of the four key parties in Maharashtra - BJP, INC, Shiv Sena and NCP, primarily the BJP politicians were active on messaging about the case. The peaks in red that we see on figure \ref{fig:cout_bihar_maha} are comprised mainly of BJP politicians.
    \item The key parties in Bihar - BJP, INC and RJD each had to address the issue briefly, especially by late-July when it appeared that this would turn into an issue in the upcoming elections.
    
    \item The peak from Bihar politicians came around the third week of July, timed alongside the trending hashtag \#NationStandsWithKangana when the narrative was spun around outsiders to Mumbai and the Mumbai film industry, including Rajput and Ranaut, being treated poorly and driven to extreme steps. Hashtags like {\#CBIForSonOfBihar} and {\#BiharsPride}, which trended during this period show how ethnonationalism became part of the meta narrative around the issue.

\end{itemize}


\subsection{Retweet analysis and media effect}

We conducted a retweet analysis to examine if tweeting about Sushant Singh Rajput or the case has an impact on the retweet rates of the individuals sending out those messages. The overwhelming share of key journalists and media houses who tweeted about the case got far higher retweets than they got on average when this was not the subject of their messaging. In this, we see the important element of public engagement, in that the press is rewarded for engaging in this story by eyeballs.

\subsubsection{Median retweets of highly followed journalists and media houses}

\begin{figure}[ht]
    \centering
    \includegraphics[width=1.0\linewidth]{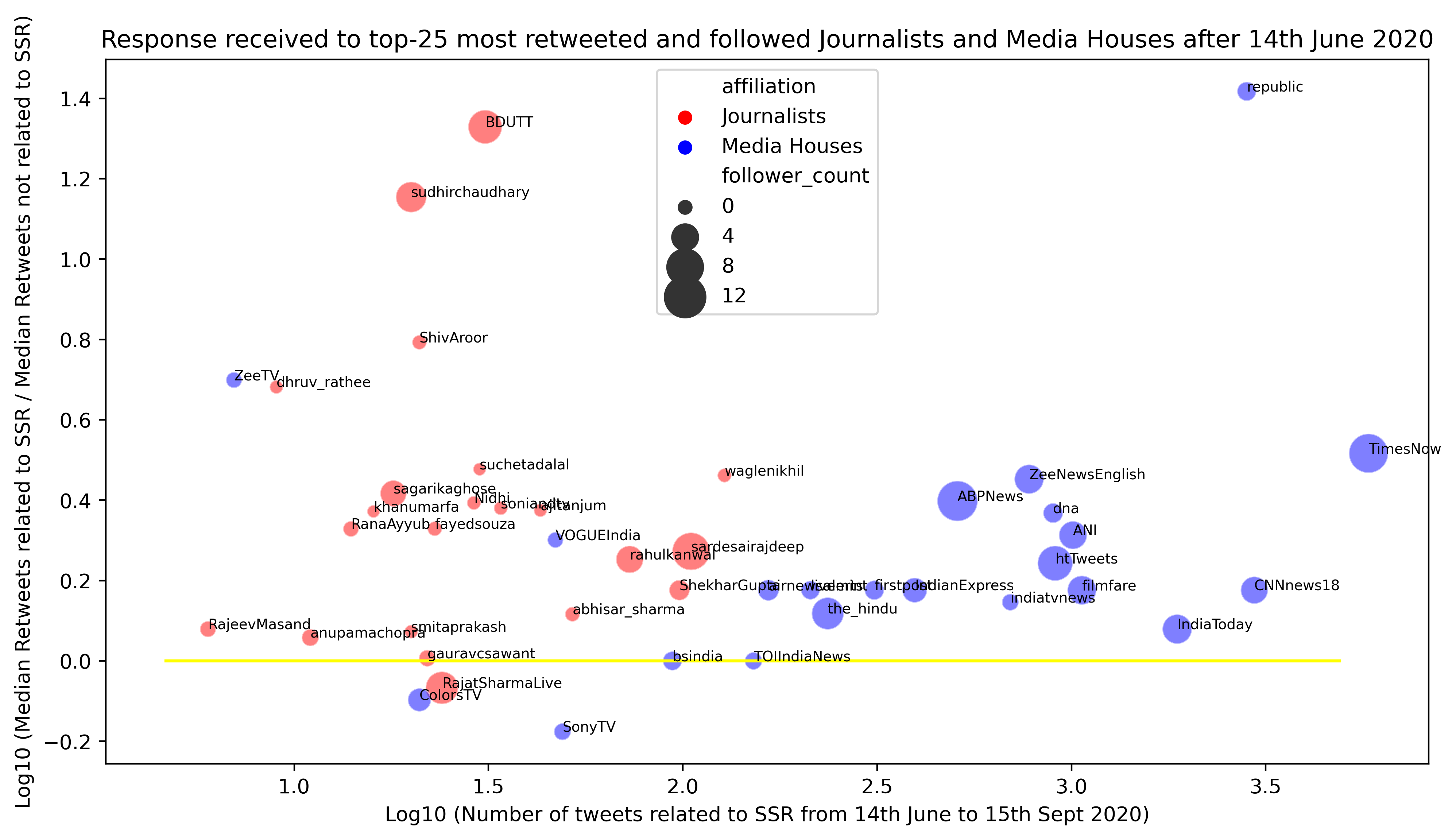}
    \setlength{\abovecaptionskip}{-0.1cm}
    \caption{Median Retweet Response of most followed Journalists and Media Houses}
    \label{fig:median_retweets}
\end{figure}

In figure \ref{fig:median_retweets}, entities above the yellow line (y>0) are termed to receive more retweets about SSR than about any other issue. The size of the bubble is proportional to the number of followers an account has.

\begin{itemize}

    \item More than 90\% of the most followed journalists and media houses receive higher median retweets on their tweets related to Sushant Singh Rajput than the tweets in which they discuss other subjects.
    \item Republic TV is the clear outlier in terms of the retweet rates it gets to its typical tweets about Sushant Singh Rajput. The channel gets a massive boost online from its followers on Twitter when it engages with the subject, several times that of most other channels. 
    \item All four major English language news channels -- Republic, TimesNow, CNNNews18, and IndiaToday have a large volume of tweets about Sushant Singh Rajput, showing that across channels there is recognition of, and tapping into the value of the story.
    \item TimesNow tweets the most about this topic.
    \item Sudhir Chaudhary and Barkha Dutt get significantly higher retweets than other journalists when they discuss this subject.
    \item Journalists like Nikhil Wagle, Rajdeep Sardesai, Shekhar Gupta, and Rahul Kanwal were among the most prolific from this set. There are, however, other less followed journalists who tweeted significantly about the topic, but are not featured in this graph which focuses on the major influencers. 
\end{itemize}

\subsubsection{Television content trend analysis}




We looked closely at the content of two specific TV channels (Republic Media Network and India Today) that are outliers in the figure \ref{fig:median_retweets} for deeper analysis of their topical choices.

Both of these TV channels have tweeted a lot about case and its related issues from their official twitter accounts, but there is a significant disparity in the kind of response and viewership they received on their social media platforms. Republic has been far more successful with the kind of content it has put out, as compared with India Today. Here are some general trends we found in the material put out by both:

\begin{itemize}
    \item Phase 1 (14th June - 13th July): India Today's content focused generally on suicide and depression strongly, whereas Republic's content had more terms and subjects around nepotism in Bollywood. Republic gave early spotlight to the points made by Kangana Ranaut, and pushed insinuations on the finances of Rhea Chakraborty's parents.
    \item Phase 2 (14th July - 13th Aug): Republic started with an interview of Kangana Ranaut which put forth much innuendo and later focused on personalities (including Ankita Lokhande, Siddharth Pithani, Shruti Modi, Mahesh Bhatt etc). It also conducted its own investigation, live on air, bringing spotlight to people around Rajput (including his cook's, gym trainer's, and key maker's interviews). India Today did not involve new entities/names so frequently into the picture, nor did it happen to conduct such investigations.
    \item Phase 3 (14th Aug - 13th Sept): India Today interviewed Rhea Chakraborty. Republic used a strategy of naming people, brought in new entities like Samuel Miranda, Gaurav Arya, Shruti Modi, etc. India Today by this point also went after naming individuals, including lesser characters who had fleeting contact with either of the key characters. Separate threads emerged out of the case - India Today created a thread for Sandalwood and Bollywood drug use angles, and the tussle between Kangana Ranaut and the BrihanMumbai Municipal Corporation (BMC).
    
    \end{itemize}

\subsection{Misinformation}

Misinformation in India does not exist in vacuum but runs with the current news cycle. It has a cause and effect relationship with the information flow in the media ecosystem \cite{panda2020topical}. Misinformation can broadly be classified into three types: disinformation which is intentionally false and designed to cause harm; misinformation which is false content but the person sharing doesn’t realise that it is false or misleading; and malinformation which is technically true information that is shared with an intent to cause harm \cite{wardle2019first}.

\begin{figure}[ht]
  \begin{subfigure}[b]{0.5\linewidth}
    \centering
    \includegraphics[height=1.0\linewidth]{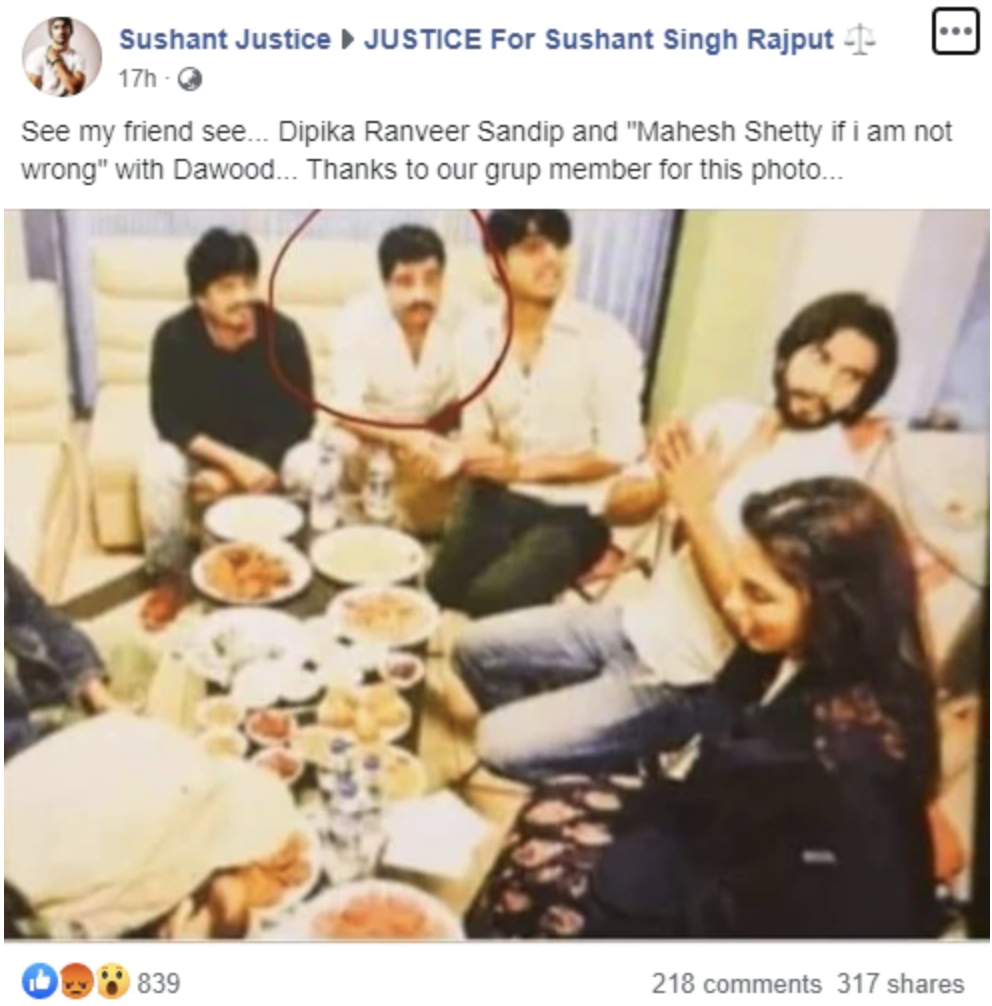}
    \caption{Fake claim linking Sushant Singh Rajput's suicide to Dawood Ibrahim (source: India Today)}
    \label{fig:Bollywood pak link)}
  \end{subfigure}
  \begin{subfigure}[b]{0.5\linewidth}
    \centering
    \includegraphics[height=1.0\linewidth]{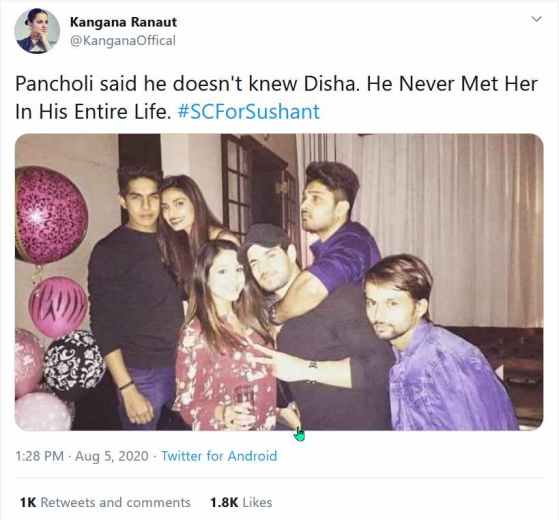}
    \caption{Fake claim suggesting link between Sushant Sing's Ex manager Disha Salian and late actress Jiah Khan's then boyfriend Sooraj Pancholi}
    \label{fig:Disha link}
  \end{subfigure}
  \caption{Some examples of misinformation stories spread on social media}
  \label{fig:misinfo}
\end{figure}

The misinformation cycle in this case is difficult to study, since there is so much innuendo and deliberate biasing of information that the lines between one kind of misleading is hard to differentiate from others. A defining feature of the news coverage, across channels, for instance is the use of provocative titles for clips on YouTube, and calling out names of the individuals in the story, which are intended to shock and clickbait. Key persons including Rhea Chakraborty, Aditya Thackeray, Disha Salian, Vinay Tiwari and Salman Khan have been targeted by misinformation, and the news cycle surrounding them makes readers more prone to click on them.

We gathered and mapped a full list of debunked misinformation collected from the Tattle archive of fake news (www.tattle.co.in) and mapped it on a dashboard that has all the stories, which is publicly available. \footnote[2]{http://joyojeet.people.si.umich.edu/ssr/} We see discernible patterns of the misinformation with the news cycle. The first is misinformation about the \textit{regular targets} who are frequently trolled on social media. The earliest debunked misinformation was claims that Rahul Gandhi referred to Rajput as a cricketer. Misinformation and jokes about Gandhi have long been successful at garnering eyeballs on WhatsApp and other platforms. Later, misinformation was about Aditya Thackeray, another frequent target of derision by trolls. The mainstream news engagement with disinformation also started early with both Aaj Tak and India.com reporting doctored tweets as Sushant Singh Rajput's last words.

One pattern we observe is of the  temporal connections of fake news to the news cycle. For instance, misinformation around the CBI, proposing that Modi authorized a probe, came about in July 2020. Later, as officers from Bihar came to probe the case, there were a series of false news circulated about them. 

Another pattern is that of fake news building off innuendo proposed by the news cycle. Here, we see the impact of Subramanian Swamy or Kangana Ranaut referring to the notion of a Bollywood mafia or need for clean ups. A particular topic that started trending that began in relation to this, was the `Bollywood-Pakistan link' (figure \ref{fig:Bollywood pak link)}). Through our mapping, we can see that this topic first trended on Twitter by Mumbai politicians and then spread across to Bihar politicians and media houses in late July. A month later in late August, a fake image of Bollywood celebrities dining with a gangster emerged. 

We also see a pattern of viral trolling that goes from one subject to another. For instance, Deepika Padukone was trolled for her post on mental health, and then included in false news about her having connections to underworld fugitive Dawood Ibrahim (figure \ref{fig:Bollywood pak link)}). More recently, Padukone has also been targeted for an investigation, based on information that came out of this case, suggesting that there are real second-order consequences of trending stories for someone  sufficiently in the public eye. 

The priming of an argument by celebrities is an important element of what makes misinformation believable. One piece of false news that came from an account claiming to be Kangana Ranaut pushing false news connecting Sooraj Pancholi (in the past, a target of a probe for the suicide death of his ex-partner) with Disha Salian, Rajput's associate who died a week before him (figure \ref{fig:Disha link}). While the image is fake, the already vilified Pancholi made for an easy, viral story.

\section{Discussion and Conclusion}

The trajectories of news coverage and misinformation around the death of Sushant Singh Rajput offer insight into the media environment in India, but also into the fractured nature of what the audience cares to consume. In a recent television interview, journalist P. Sainath rued that scores of suicides of farmers, driven to desperation by poverty and state failure, had failed to garner even a small percentage of the attention that this case had received. That the audience has consistently rewarded news channels for following this story, for instance, through meteoric ratings for the Republic news network which has offered the most aggressive coverage, are testament to the citizenry's complicity.

Undoubtedly, the specifics of the story, particularly Rajput's journey as an outsider, breaking into the difficult world of show business with its well-known nepotism is an important part of why it had such affective value. The notion of Rajput as an outsider is not unlike past, successful narratives of other outsiders who have taken on a system - most prominently that of Narendra Modi taking on the entrenched Congress institutions. Empathizing with Rajput is rooting for the underdog.

The data show an important role played by politicians, especially the BJP, in proposing a 'murder' alternative to the 'suicide' narrative. There was a real opportunity to address mental health and depression early in news cycle, but the stories quickly devolved to allusive concoctions. The move towards conspiracies was accompanied by multiple supporting actors - the local police was proposed as incompetent, or in cahoots with the cabal, the state government itself was presented as nepotistic and inimical to the interests of poor outsiders. As research into online groups claiming to seek justice for Sushant Singh Rajput have shown, a heady mix of ultranationalism, casteism, distrust of Muslims, and misogyny come are drivers of some of the online action that we have seen in recent months.\footnote[3]{https://www.newslaundry.com/inside-the-online-cult-of-justice-for-SSR} And as much recent work into political speech has also shown, an ecosystem of outrage exists in place in India, and it may be far from coincidental that a lot of the celebrities being trolled in the aftermath of the suicide were among those who were critical of the government in the past.

The array of stakeholders, each with their own interests in moving the story forward, are an important part of what has driven this to the point of being perhaps the top national story, despite being in the middle of the worst pandemic and economic crisis India has known. The timing of the suicide, in the midst of the COVID-19 crisis and lockdown with many urban middle-class Indians stuck at home, probably had a role in driving up purchase for the story. The Sushant Singh Rajput case came at a point when the prime time for a while had been saturated simply with news of the virus. The story offered a diversion. While the massive win for the actor's posthumously released film offered a silver lining, the creation of villains gave people an object for vengeance.

Several public figures, long outside the spotlight in the midst of the COVID-19 crisis found their way back into the news. The media channels that amplified these stories - particularly the Republic and Times of India groups, have a financial incentive to get more viewers. The political parties that engaged in the story arguably benefited from being in the spotlight and used it as a means to attack their opponents. Perhaps the most interesting case is that of sitting parliamentarian Subramanian Swamy, once a central player in the policy circles, now wilfully mongering deceit, and also the most popular Twitter handle in wordclouds of trolls. 

Finally, there is also an unmistakable gendered aspect to this. Several high profile female stars have taken their own lives over the years, and many of these stories have ended with victim blaming or stayed in the news cycle for very short periods of time. The victims of this case have mostly been women -- Rhea Chakraborty was slandered and hounded, eventually ending up in jail without bail. As the scandal devolved into allegations about drugs - the Indian Narcotics Control Bureau (NCB) summoned four movie actors - Deepika Padukone, Sara Ali Khan, Shraddha Kapoor, and Rakul Preet Singh, all women, and at the time of this writing, all intensely targeted by trolling and mainstream media speculation online.

The Sushant Singh Rajput case is a story of one man's personal journey that ends in tragedy. The events that followed may tempt us to think that this offers a window into the ways that online culture has changed society and media in India. But the truth may be more chilling than that. While social media may have facilitated certain kinds of virality and speed with which the narratives have changed, the case and its victims are a reminder of ways the patriarchy is alive and well, and always readying its blades for the next execution.

\bibliographystyle{ACM-Reference-Format}
\bibliography{main}

\end{document}